\documentclass[12pt,twoside,pointlessnumbers,smallheadings]{article}
\usepackage{graphicx}
\input psfig.sty

\newcommand{\ag}{a_{3g}}
\newcommand{\BR}{{\cal B}}

\newcommand{\psp}{\psi^{\prime}}
\newcommand{\psip}{\psi(2S)}
\newcommand{\pspp}{\psi^{\prime \prime}}
\newcommand{\psipp}{\psi(3770)}
\newcommand{\jpsi}{J/\psi}

\newcommand{\EE}{e^+e^-}

\newcommand{\kskl}{K^0_SK^0_L}
\newcommand{\KKSC}{K^{*+}K^-}
\newcommand{\KKSN}{K^{*0}\overline{K^0}}

\newcommand{\OP}{\omega\pi^0}

\newcommand{\rhopi}{\rho\pi}

\newcommand{\ra}{\rightarrow}

\newcommand{\beq}{\begin{equation}}
\newcommand{\eeq}{\end{equation}}
\newcommand{\beqn}{\begin{eqnarray}}
\newcommand{\eeqn}{\end{eqnarray}}
\newcommand{\beqns}{\begin{eqnarray*}}
\newcommand{\eeqns}{\end{eqnarray*}}
\newcommand{\bfg}{\begin{figure}}
\newcommand{\efg}{\end{figure}}
\newcommand{\bitm}{\begin{itemize}}
\newcommand{\eitm}{\end{itemize}}
\newcommand{\bnum}{\begin{enumerate}}
\newcommand{\enum}{\end{enumerate}}
\newcommand{\btbl}{\begin{table}}
\newcommand{\etbl}{\end{table}}
\newcommand{\btbu}{\begin{tabular}}
\newcommand{\etbu}{\end{tabular}}

\newsavebox{\arrect}
\savebox{\arrect}(0,0){%
\setlength{\unitlength}{0.5cm}
\thicklines
\put(-4,1){\line(1,0){8.0}}
\put(4,1){\line(0,-1){2.0}}
\put(4,-1){\line(-1,0){8.0}}
\put(-4,-1){\line(0,1){2.0}}
\put(0,-1){\vector(0,-1){1.0}}
}

\newsavebox{\arrhomb}
\savebox{\arrhomb}(0,0){%
\setlength{\unitlength}{1cm}
\thicklines
\put(0,0.5){\line(-2,-1){1.0}}
\put(0,0.5){\line(2,-1){1.0}}
\put(0,-0.5){\line(-2,1){1.0}}
\put(0,-0.5){\line(2,1){1.0}}
\put(0,-0.5){\vector(0,-1){0.5}}
\put(-0.15,-0.7){\makebox(0,0)[r]{no}}
\put(1.0,0.0){\vector(1,0){2.0}}
\put(2.0,0.05){\makebox(0,0)[b]{yes}}
}

\newsavebox{\arrparall}
\savebox{\arrparall}(0,0){%
\setlength{\unitlength}{0.5cm}
\thicklines
\put(-3,1){\line(1,0){8.0}}
\put(5,1){\line(-1,-1){2.0}}
\put(3,-1){\line(-1,0){8.0}}
\put(-5,-1){\line(1,1){2.0}}
\put(0,-1){\vector(0,-1){1.0}}
}

\begin{document}          

\begin{titlepage}
\title{Universal phase between strong and EM interactions 
\footnote{work in collaboration with C.~Z.~Yuan and X.~H.~Mo}}

\author{Ping Wang~\footnote{E-mail:wangp@IHEP.ac.cn}
\\
{\small  Institute of High Energy Physics, 
Beijing 100039, China  } }

\date{\mbox{}}
\maketitle

\begin{abstract}
It is shown that the experimental data of $\psi^\prime$ and
$\psi^{\prime\prime}$ are consistent with a $-90^\circ$ phase
between the strong and eletromagnetic decay amplitudes. 
The $e^+e^-\rightarrow \rho\pi$ measured at 
$\psi^{\prime\prime}$ 
is also consistent with the branching ratio predicted by 
Rosner's scenario on $\rho\pi$ puzzle in charmonium physics. 
This scenario leads to a possible large charmless branching ratio  
in $\psi^{\prime\prime}$ decays.

\end{abstract}

\end{titlepage}

\section{Motivations}

It has been known from experimental data that in two-body 
$\jpsi$ decays, the relative phase between the 
strong decay amplitude $a_{3g}$ and electromagnetic (EM) decay
amplitude $a_\gamma$ is orthogonal for the decay modes 
$1^+0^-$ ($90^\circ$)~\cite{suzuki}, 
$1^-0^-$  $(106 \pm 10)^\circ$~\cite{jousset}, 
$0^-0^-$ $(89.6 \pm 9.9)^\circ$~\cite{suzuki2}, 
$1^-1^-$ $(138 \pm 37)^\circ$~\cite{kopke} and 
$N\overline{N}$ $(89 \pm 15)^\circ$~\cite{baldini}. 

It was argued~\cite{gerard} that this large phase follows from  
the orthogonality of three-gluon and one-photon virtual processes.
The question arises:
is this phase universal for quarkonium decays?
How about $\psp$, $\pspp$ and $\Upsilon(nS)$ decays?

\section{The phase between strong and EM amplitudes 
in $\psp$ decays}

\begin{figure}[hbt]
\begin{minipage}{5cm}
\centering
\includegraphics[height=3.5cm,width=4.0cm,angle=0]{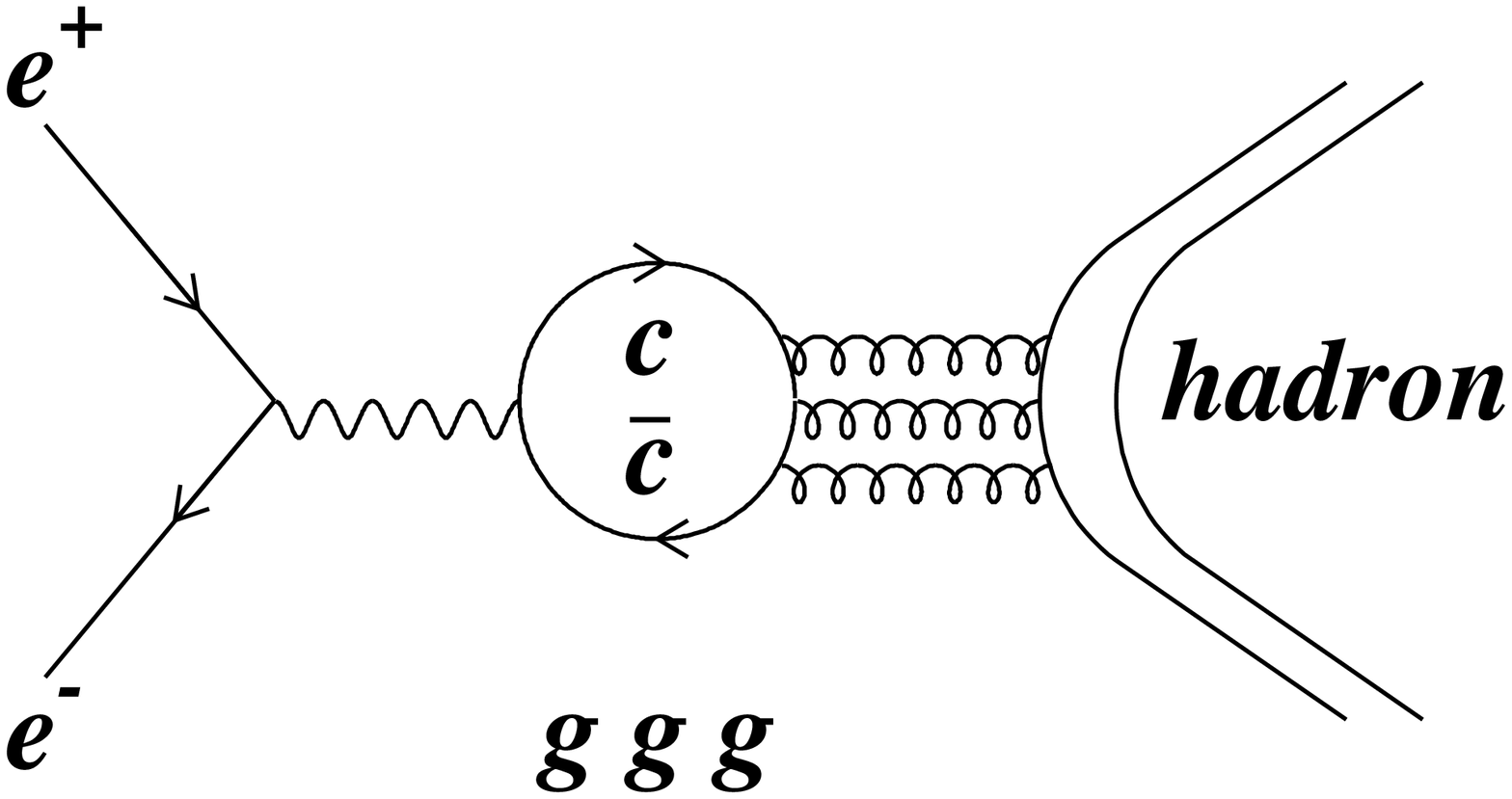}
\caption{\label{ggg}
strong decay}
\end{minipage}
\begin{minipage}{5cm}
\centering
\includegraphics[height=3.5cm,width=4.0cm,angle=0]{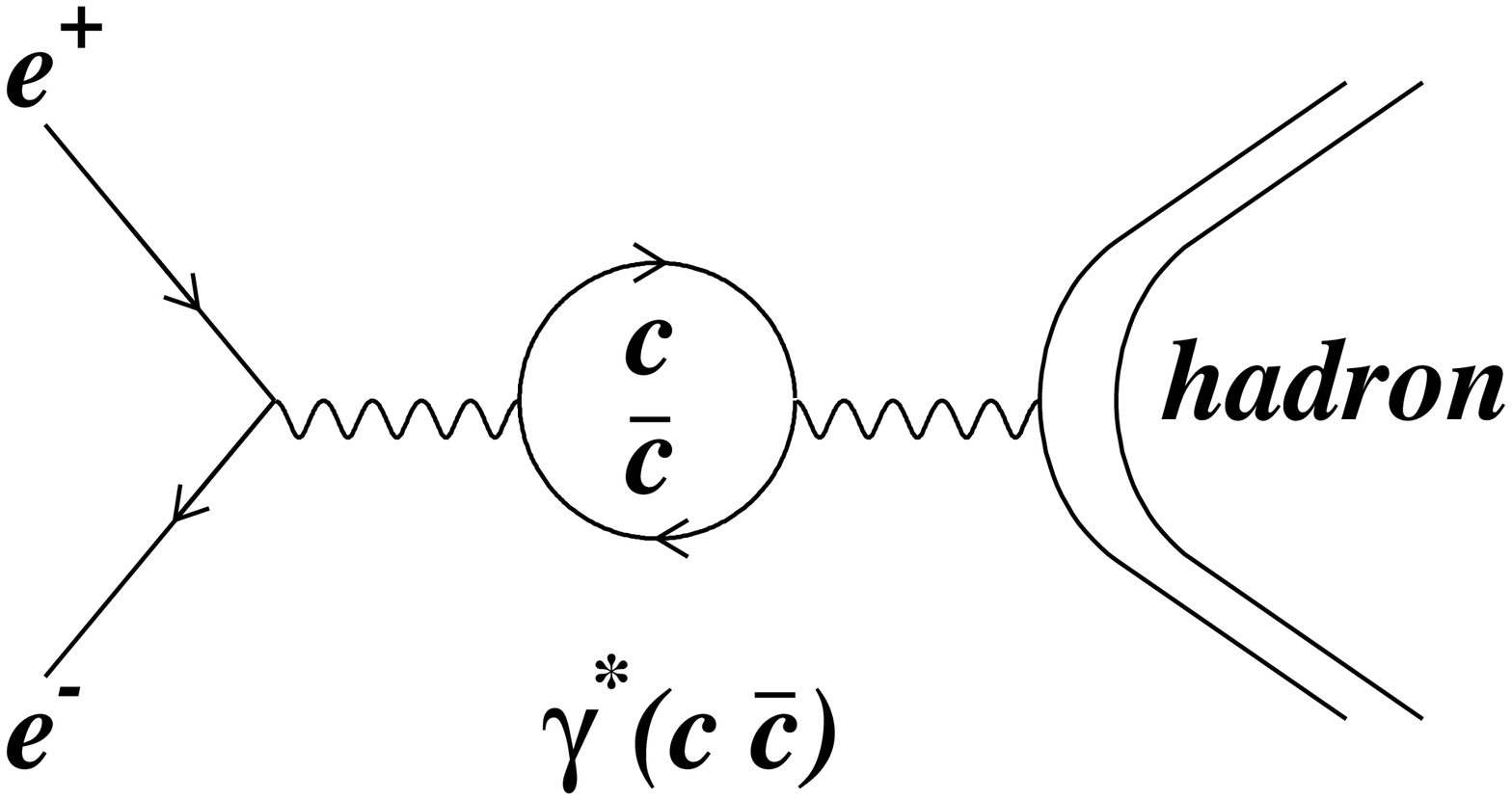}
\caption{\label{gcc}
EM decay}
\end{minipage}
\begin{minipage}{5cm}
\centering
\includegraphics[height=3.5cm,width=4.0cm,angle=0]{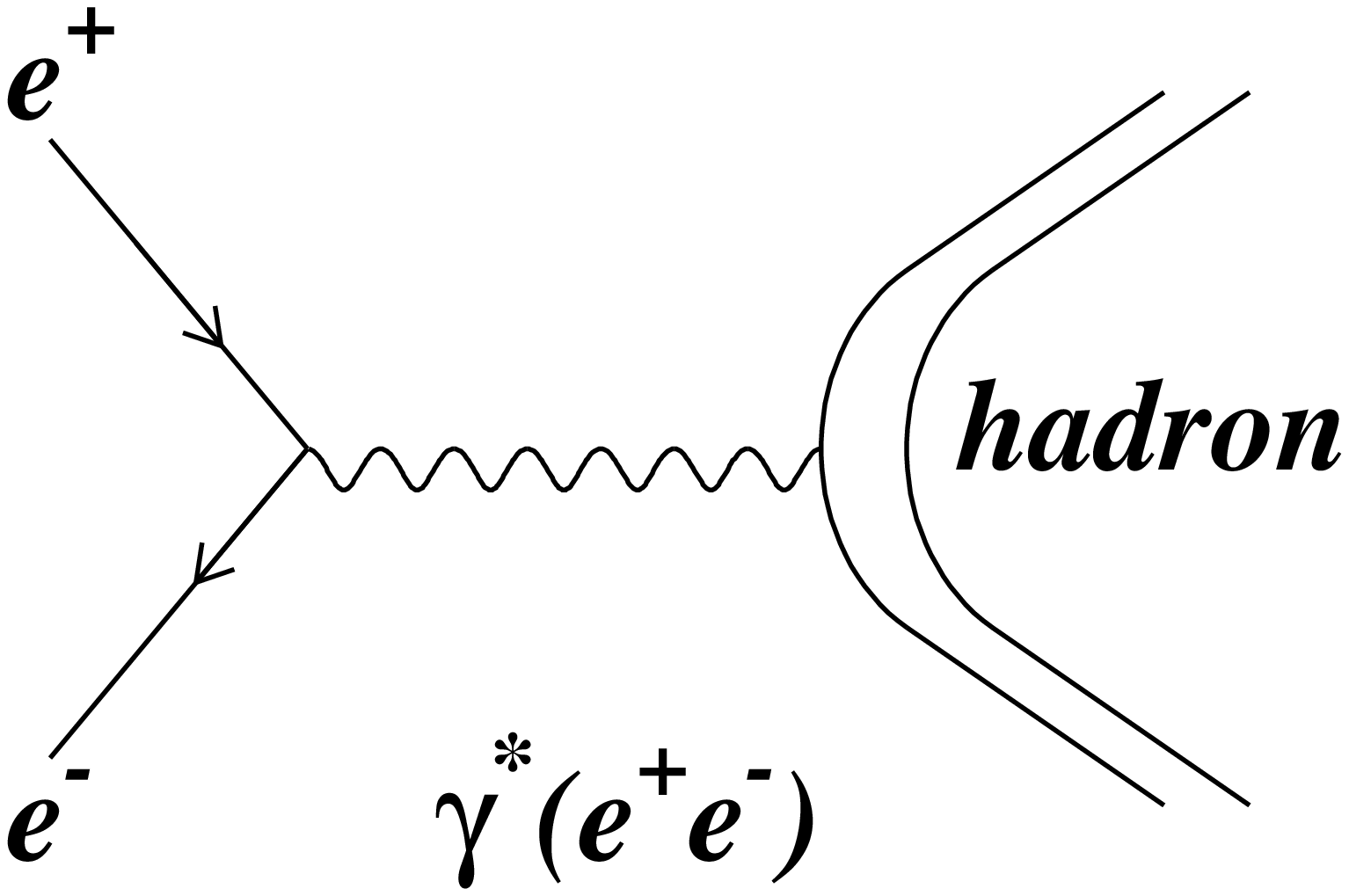}
\caption{\label{gee}
continuum}
\end{minipage}
\end{figure}

Recently, more $\psp$ data has been available.
Most of the branching ratios are measured in $\EE$ colliding
experiments. For these experiments, 
there are three diagrams~\cite{rudaz,interf} 
which contribute to the processes as shown in 
Fig.~(\ref{ggg},\ref{gcc},\ref{gee}).

Until recently, the diagram
in Fig.~(\ref{gee}) has been neglected in the analysis of
$\psp$ decays. But it leads to a continuum cross section and more
important, it interferes with the amplitude of 
Fig.~(\ref{ggg}). So it
affects the measured branching ratios
significantly and alters the determination of the
phase~\cite{interf}.  

For the $\EE\rightarrow1^-0^-$ processes, the amplitudes depend on the
three diagrams in the way~\cite{haber}: 
\beq
\begin{array}{ccl}
A_{\OP} &=& 3(a_\gamma+a_c)~~ , \\
A_{\rhopi} &=& \ag+a_\gamma+a_c~~ , \\
A_{\KKSC} &=& \ag+\epsilon+a_\gamma+a_c~~ , \\
A_{\KKSN} &=& \ag+\epsilon-2(a_\gamma+a_c) ~~.
\end{array}
\label{arp}
\eeq
where $\epsilon$ is the SU(3) symmetry breaking parameter.  
They can then be expressed as 
\beq
\begin{array}{ccl}
A_{\OP}   &=& [1 + B(s)] \cdot {\cal F}_{\OP}(s) ~~, \\
A_{\rhopi}&=& [({\cal C}e^{i\phi}+1)B(s)+1 ] \cdot {\cal F}_{\OP}(s)/3~~, \\
A_{\KKSC} &=& [({\cal C} {\cal R} e^{i\phi}+1)B(s)+1 ] 
                                             \cdot {\cal F}_{\OP}(s)/3~~, \\ 
A_{\KKSN} &=& [({\cal C} {\cal R} e^{i\phi}-2)B(s)-2 ] 
                                             \cdot {\cal F}_{\OP}(s)/3~~. \\
\end{array} 
\label{arpnew}
\eeq
where 
${\cal R}=|(a_{3g}+\epsilon)/a_{3g}|$, $C=|a_{3g}/a_\gamma|$, and 
$$B(s) \equiv \frac{3\sqrt{s}\Gamma_{ee}/\alpha}{s-M^2+iM\Gamma_t}~~.$$
On top of the resonance, $B(s)=-i3B_{ee}/\alpha$ with phase of
$-90^\circ$. If $\phi$ which is the phase between $a_{3g}$
and $a_\gamma$ is $-90^\circ$, then the relative phase between $a_{3g}$
and $a_c$ is $180^\circ$ for $\rhopi$ and $\KKSC$, but $0^\circ$ for 
$\KKSN$. The interference pattern due to this phase explains the small signal
of $\rhopi$ and $\KKSC$ but large signal of $\KKSN$ observed by BES and
CLEOc at $\psp$~\cite{zhu,cleoc}. We suggest that in $\psp \ra VP$
decays, the strong and EM amplitudes are still orthogonal and the sign
of the phase must be negative~\cite{possiblephase}.    

For $\psp \ra PP$ decays, the calculation~\cite{pspp} compared
with the BES measurement of $B(\psp \ra \kskl)$~\cite{beskskl}, 
leads to the conclusion that
the phase between strong and EM amplitudes is either 
$(-82 \pm 29)^\circ$ or $(121 \pm 27)^\circ$. 

\section{$\pspp \ra \rhopi$ and Rosner's scenario on $\rhopi$ puzzle}%

As we turn to such phase in $\pspp$ decays, we get an extra prize
which is the solution of the long-lasting $\rhopi$ puzzle in 
charmonium decays. First we must digress to Rosner's scenario. 

While $\rhopi$ has the largest branching ratio among the 
hadronic final states in $\jpsi$ decays, 
the same mode was not found in $\psp$ decays 
for a long time (recently, BES and
CLEOc report its branching ratio at the order 
of $10^{-5}$~\cite{zhu,cleoc}).  Rosner proposed that this is due to
the mixing between $\psi(2^3S_1)$ and $\psi(1^3D_1)$ states~\cite{rosner}:
\beqn
\langle\rhopi |\psp\rangle =\langle \rhopi | 2^3 S_1 \rangle \cos \theta
                  -\langle \rhopi | 1^3 D_1 \rangle \sin \theta~, \nonumber \\ 
\langle\rhopi |\pspp\rangle=\langle \rhopi | 2^3 S_1 \rangle \sin \theta
                  +\langle \rhopi | 1^3 D_1 \rangle \cos \theta~, \nonumber
\eeqn
where $\theta=12^\circ$ 
is the mixing angle~\cite{rosner}. The missing of $\rhopi$ 
in $\psp$ decay is due to the cancellation 
of the two terms in  $\langle\rhopi |\psp\rangle$. This scenario is 
simple, and it predicts with little uncertainty that 
${\cal B}_{\pspp\rightarrow\rhopi}=(6.8\pm2.3)\times10^{-4}~~,
$ or
\beq
\sigma^{Born}_{\EE \ra \pspp \ra \rhopi} = (7.9 \pm 2.7) \mbox{pb}
\label{sipp2rhopi}
\eeq
with BES latest result on $\BR(\jpsi \ra \rhopi)$~\cite{besjpsi}. 

On the other hand, using CLEOc measurement of $\EE \ra \rhopi$ at
3.67GeV~\cite{cleoc}, scaled to 3.77GeV according to $1/s^2$, we obtain
\beq
\sigma^{Born}_{\EE \ra \gamma^* \ra \rhopi} 
(3.770GeV) = (7.5 \pm 1.8) \mbox{pb}.
\label{ee2rhopi}
\eeq 
The Born cross sections in Eqs.(\ref{sipp2rhopi}) and (\ref{ee2rhopi}) are
comparable. The question arises: how do they interfere?

\begin{figure}[hbt]
\begin{minipage}{15.0cm}
\centering
\includegraphics[height=10.0cm,width=15.0cm,angle=0]{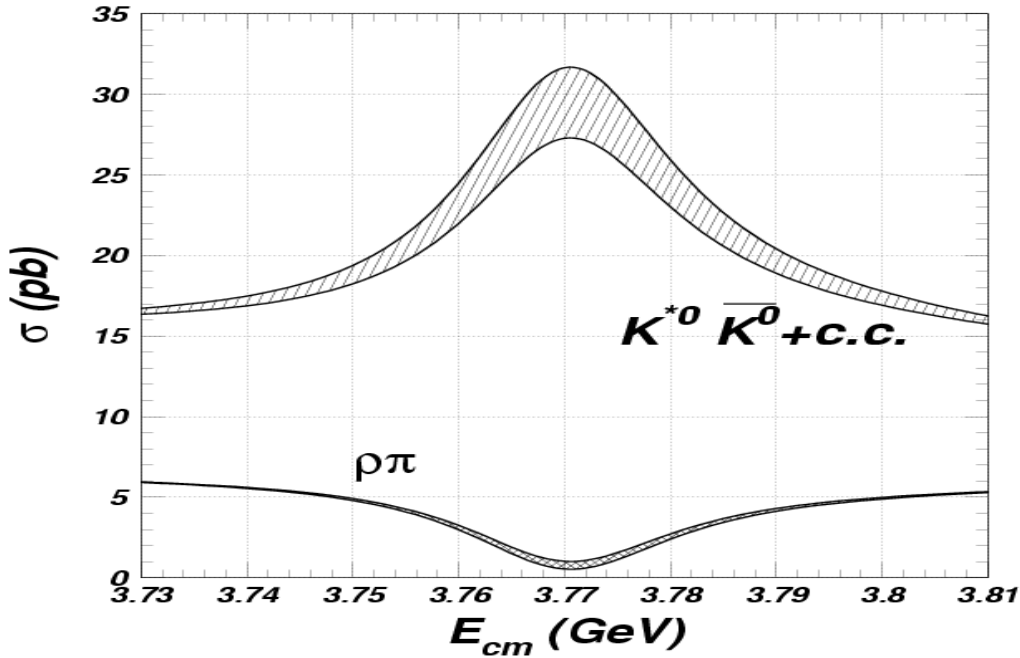}
\caption{\label{rpkk}
The $\EE \ra \rhopi$ and $\EE \ra \KKSN + c.c.$ cross sections 
around $\pspp$ peak, assuming Rosner's scenario and $-90^\circ$ phase
between strong and EM amplitudes. Hatched area is due to 
an unknown phase between the $2^3S_1$ and $1^3D_1$ matrix elements.}
\end{minipage}
\end{figure}

As a matter of fact, MARK-III measured this cross section
at $\pspp$ peak, and gave~\cite{zhuyn}
\beq
\sigma_{\EE \ra \rhopi}(3.770GeV) < 6.3 \mbox{pb},
\eeq
which is already smaller than the continuum cross section in
Eq.(\ref{ee2rhopi}). We expect BES and CLEOc to bring this
value further down. This means~\cite{sipp2ropi}:
\bitm
\item{There must be destructive interference between resonance and
  continum, i.e. the phase between the strong and EM amplitudes is
  again $-90^\circ$.}
\item {$\BR(\pspp \ra \rhopi) \approx (6 \sim 7) \times 10^{-4}$,
  i.e. Rosner's scenario gives correct prediction!}
\eitm
If we scan $\pspp$, we shall find the cross sections of $\EE \ra
\rhopi$ and $\EE \ra \KKSN + c.c.$ versus energy like the curves
in Fig.(\ref{rpkk}). In the figure, the hatched area is due to 
an unknown phase between the $2^3S_1$ and $1^3D_1$ matrix 
elements~\cite{sipp2ropi}. The $\KKSC + c.c.$ cross section
is similar to $\rhopi$.

\section{The phase in $\Upsilon$ decays}
CLEO observed $K^*K$ but not $\rhopi$ in $\Upsilon$
decays~\cite{upsilon}. It can 
be due to the same interference pattern. We suppose 
the $K^*K$ signal in CLEO observation is mainly $\KKSN$, not $\KKSC$.    

\section{Rosner's scenario and enhanced modes in $\psp$ decays}

Recently BES found modes which are enhanced in $\psp$ decays 
relative to $\jpsi$. One
of them is $\kskl$: $\BR(\jpsi
\ra\kskl)=(1.82\pm0.04\pm0.13)\times10^{-4}$ and 
$\BR(\psip \ra \kskl)=(5.24\pm0.47\pm0.48)\times10^{-5}$ with
$Q_h=(28.8\pm3.7) \%$ versus 12\% rule. If such enhancement is 
due to the mixing of $2^3S_1$ and $1^3D_1$ states, then
we expect~\cite{pspp2kskl} 
$(1.2 \pm 0.7)\times10^{-6} \le 
\BR(\psipp \ra \kskl) \le (3.8 \pm 1.1)\times~10^{-5}$.
Here the range is due to an unknown phase between 
$\langle \kskl | 2^3S_1 \rangle$ and $\langle \kskl | 1^3D_1
\rangle$. If this phase is 0, then the prediction is at the upper
bound. 

Currently BES gives an upper limit~\cite{besppkskl} 
$\BR(\pspp \ra \kskl) < 2.1 \times 10^{-4}$. We expect CLEOc to give the
branching ratio. 

\section{$\pspp$ decays to charmless final states}

It has been noticed that there is hadronic excess in $\psp$ decays
which has no parallel in $\Upsilon$ physics~\cite{suzuki,gu}:
\beq
Q_1=\frac{\BR(\psp \ra ggg + \gamma gg)}{\BR(\jpsi \ra ggg + \gamma gg)} 
= (26.0 \pm 3.5)\%~~
\label{qvone}
\eeq
versus 12\% rule. It indicates that most of the $\psp$ partial widths 
via gluons go to the
final states which are enhanced in $\psp$ decays. Now we do not know
what these final states are. 
The question arises: what is their branching ratio in
$\pspp$ decays? There has been experimental indication that $\pspp$ 
has a substantial charmless branching ratio, although it comes 
with large uncertainties. This was addressed again
recently~\cite{rosner04}. So let us estimate the possible
combined branching ratio of these final states in $\pspp$ decays.    

We define the suppression and enhancement
factor~\cite{rosner04} 
\beq Q(f) \equiv
\frac{\Gamma(\psp \ra f) }{\Gamma(\jpsi \ra f)}
         \frac{\Gamma(\jpsi \ra \EE)}{\Gamma(\psp \ra \EE)}~.
\label{definec} 
\eeq 
$Q(f)<1$ means the final state $f$ is suppressed in $\psp$ decays
relative to $\jpsi$; 
$Q(f)>1$ means it is enhanced; $Q(f)=1$ means it observes the 12\%
rule. 
 
\begin{figure}[hbt]
\begin{minipage}{15cm}
\centering
\includegraphics[height=10cm,width=15cm,angle=0]{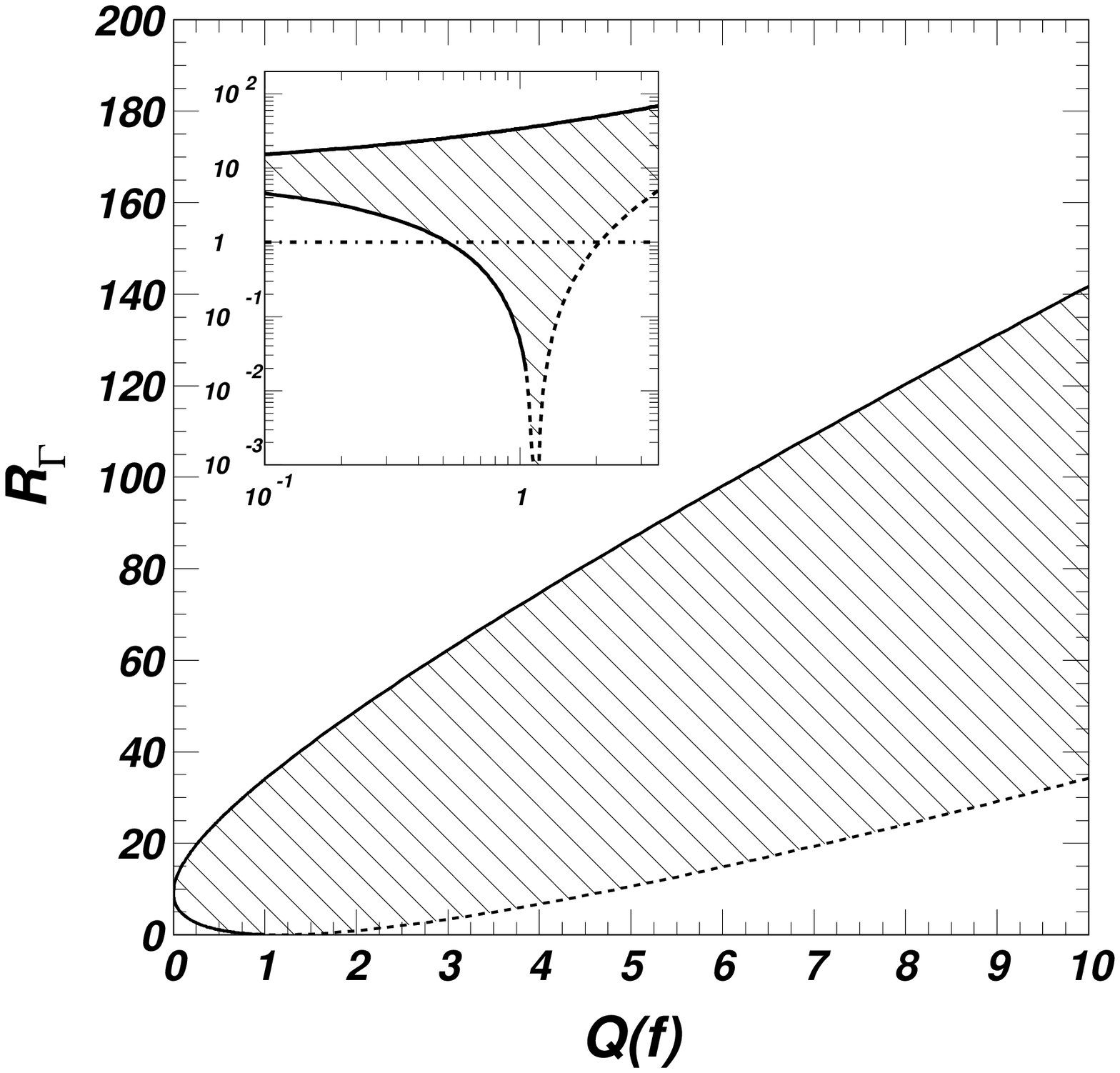}
\caption{\label{fig:enhsup}
$R_\Gamma$ as a function of $Q(f)$. The solid
contour corresponds to no extra phase between the matrix elements
$\langle f | 2^3S_1 \rangle$ and $\langle f | 1^3D_1 \rangle$; 
dashed contour corresponds to a relative
negative sign between the matrix elements; 
the hatched area corresponds to other non-zero phase between 
the marix elements.}
\end{minipage}
\end{figure}

In the $2S-1D$ mixing scheme, for any final state, its partial
width in $\pspp$ decay can be related to its partial widths in 
$\jpsi$ and $\psp$ decay with an unknown parameter which is the relative
phase between the matrix elements $\langle f | 2^3S_1 \rangle$ 
and $\langle f | 1^3D_1 \rangle$. This unknown phase constrains the 
predicted $\Gamma(\pspp \ra f)$ in a finite range. 
We calculate $R_\Gamma \equiv \Gamma(\pspp \ra f) / \Gamma(\jpsi \ra
f)$ as a function of $Q(f)$ and plot it in Fig.~(\ref{fig:enhsup}).  
In the figure the solid
contour corresponds to the solution with no extra phase between 
$\langle f | 2^3S_1 \rangle$ and $\langle f | 1^3D_1 \rangle$; dashed
contour corresponds to the solution with a relative
negative sign between $\langle f | 2^3S_1 \rangle$ and 
$\langle f | 1^3D_1 \rangle$; the hatched area corresponds to the
solution with other non-zero phase between 
$\langle f | 2^3S_1 \rangle$ and $\langle f | 1^3D_1 \rangle$.
From Fig.~(\ref{fig:enhsup}) we see that those final states with large
$Q(f)$ may contribute a combined large branching ratio in $\pspp$
decays.  

The decays of $\jpsi$ and $\psp$ are classified into gluonic decays
($ggg$), electromagnetic decays ($\gamma^*$), radiatve decays into
light hadrons ($\gamma gg$), and OZI allowed decays into lower mass 
charmonium states. By subtracting the second to fourth classes, we
obtain $\BR(\jpsi \ra ggg) \approx (69.2 \pm 0.6)\%$ and 
$\BR(\psp \ra ggg) \approx (18.0 \pm 2.4)\%$.
Among these final states, we know that VP and VT final states 
have $Q(f)<1$, and $N\bar{N}$ have $Q(f) \approx 1$.
Together they consist 5.4\% of $\jpsi$ decays
and $1.8 \times 10^{-3}$ of $\psp$ decays.  
We subtract their branching ratios from the total branching
ratio of gluonic decays of $\jpsi$ and $\psp$. The remaining  
63.8\% of $\jpsi$ decay and 17.8\% of $\psp$ decay 
which go to final states through $ggg$ either have $Q(f)>1$ or
$Q(f)$ unknown. On the average these final states have $Q(rem) \approx
2.19$. For this $Q$ value, the maximum $R_\Gamma$ is 51.6. So the maximum
partial width of these final states in $\pspp$ is 
$\Gamma_{tot}(\jpsi) \times 63.8\% \times 51.6$ 
which is 3.0MeV, or 13\% of the total $\pspp$ decay. 

The above maximum value of $R_\Gamma$ comes if there is no extra phase
between  $\langle f | 2^3S_1 \rangle$ 
and $\langle f | 1^3D_1 \rangle$. There are reasons to assume that this
is the case: (1) in the matrix element of $\langle \rhopi | \psp \rangle$,
there is almost complete cancellation between 
the contributions from $2^3S_1$ and $1^3D_1$
matrix elements, so the phase between them must close to 0; (2) if
the phase between the strong and EM ampitudes is universal,
then there is no extra phase between  $2^3S_1$ and $1^3D_1$
matrix elements due to strong interactions, since there is no extra
phase between the two matrix elements due to EM interactions, as in
the calculations of leptonic decays. So we suppose that the partial
widths of these final states are at the maximum values calculated
here.  
 
The calculations here take the averaged $Q(f)$ so serve as a 
rough estimation. The exact charmless partial width
should be the sum of individual final states which in general have 
different values of $Q(f)$. But at present, experiments do 
not provide enough informationm to conduct such calculation. 
Nevertheless, the calculation here shows that a large charmless 
branching ratio in $\pspp$ decays, e.g. more than 10\%, is not a
surprise. It is well explained in the $2S-1D$ mixing scenario. 
Measuring the charmless branching ratio of $\pspp$ decays, 
both inclusive and exclusive, should be a primary physics goal 
for BES and CLEOc. 

\section{Summary}
The $\psp \rightarrow 1^-0^-$ and $0^-0^-$ data collected in $e^+e^-$
experiments are consistent with a $-90^\circ$ phase between 
strong and electromagnetic interactions. 
This phase also holds in OZI suppressed decays of $\pspp$. This
is from the measured $\rho\pi$ cross sections at $\pspp$ and
3.67GeV. At the same time these measurements give  
$\BR(\pspp \rightarrow \rho\pi)$ which agrees with the 
prediction by Rosner in his scenario explaining the
$\rho\pi$ puzzle. This scenario would be further supported if the large
charmless branching ratio in $\pspp$ decays is confirmed by
experiments.

\section*{Acknowledgments}

This work is supported in part by the 100 Talents Program of CAS
under Contract No. U-25.

\end{document}